\documentclass[prd,aps]{revtex4}

\usepackage{graphicx,subfigure}

\newcommand{\be}{\begin{equation}}
\newcommand{\ee}{\end{equation}}
\newcommand{\bea}{\begin{eqnarray}}
\newcommand{\eea}{\end{eqnarray}}
\newcommand{\bd}{\begin{displaymath}}
\newcommand{\ed}{\end{displaymath}}

\newcommand{\q}{ 1+ (1-q)t }
\newcommand{\qq}{( 1+ (1-q)t )^{\frac{1}{1-q}}}
\newcommand{\qi}{ \frac{1}{1-q} }
\newcommand{\qs}{ \frac{s}{1-q} }

\newcommand{\la}{ \ln_q t  }
\newcommand{\x}{ t^{1-q}  }

\newcommand{\op}{ \oplus  }
\newcommand{\om}{ \ominus  }
\newcommand{\ot}{ \otimes  }
\newcommand{\od}{ \oslash }

\newcommand{\lb }{ \left( }
\newcommand{\rb }{ \right ) }

\begin{document}


\title{
On the $q$-Laplace transform in the non-extensive statistical physics  }

\author{ Won Sang Chung }
\email{mimip4444@hanmail.net}

\affiliation{
Department of Physics and Research Institute of Natural Science, College of Natural Science, Gyeongsang National University, Jinju 660-701, Korea
}

\date{\today}

\begin{abstract}
In this paper, q-Laplace transforms related to the non-extensive thermodynamics are investigated by using the algebraic operation of the non-extensive calculus. The deformed simple harmonic problem is discussed by using the q-Laplace transform.

\end{abstract}

\maketitle

\section{Introduction}

Boltzman-Gibbs statistical mechanics shows how fast microscopic physics with short-range interaction has as effect on much larger space-time scale. The Boltzman-Gibbs entropy is given by
\be
S_{BG} = - k \sum_{i=1}^W p_i \ln p_i = k \sum_{i=1}^W  \ln \frac{1}{p_i}
\ee
where $k$ is a Boltzman constant, $W$ is a total number of microscopic possibilities of the system and $p_i$ is a probability of a given microstate among $W$ different ones satisfying $ \sum_{i=1}^W p_i = 1 $. When $ p_1 = \frac{1}{W} $, we have $S_{BG} = k \ln W $.

Boltzman-Gibbs theory is not adequate for various complex, natural, artificial and social system. For instance, this theory does not explain the case that a zero maximal Lyapunov exponent appears. Typically, such situations are governed by power-laws instead of exponential distributions. In order to deal with such systems, the non-extensive statistical mechanics is proposed by C.Tsallis [1,2]. The non-extensive entropy is defined by
\be
 S_q = k ( \Sigma_{i}^W p_i^q - 1 ) / ( 1-q )
\ee
The non-extensive entropy has attracted much interest among the physicist, chemist and mathematicians who study the thermodynamics of complex system [3]. When the deformation parameter $q$ goes to 1, Tsallis entropy (2) reduces to the ordinary one (1). The non-extensive statistical mechanics has been treated along three lines:

\vspace{0.5cm}

1. Mathematical development [4, 5, 6]

2. Observation of experimental behavior [7]

3. Theoretical physics ( or chemistry) development [8]

\vspace{0.5cm}

The basis of the non-extensive statistical mechanics is q -deformed exponential and logarithmic function which is different from those of Jackson's [9]. The q-deformed exponential and q-logarithm of non-extensive statistical mechanics is defined by [10]
\be
\la  = \frac{ \x -1 } {1-q}, ~~~ ( t >0 )
\ee
\be
e_q (t)  =  \qq , ~~~ (x, q \in R )
\ee
where $ \q >0 $.

From the definition of q-exponential and q-logarithm, q-sum, q-difference, q-product and q-ratio are defined by [ 5, 6]
\bea
x \op y &=& x + y + (1-q) xy \cr
x \om y &=& \frac{ x-y}{1+(1-q)y} \cr
x \ot y &=& [  x^{1-q}+ y^{1-q} -1 ]^{\qi} \cr
x \od y &=& [  x^{1-q} - y^{1-q} +1 ]^{\qi}
\eea
It can be easily checked that the operation $\op$ and $ \ot $ satisfy commutativity and associativity. For the operator $ \op$, the identity additive is $0$, while for the operator $\ot$ the identity multiplicative is $1$. Indeed, there exist an analogy between this algebraic system and the role of hyperbolic space in metric topology [11].  Two distinct mathematical tools appears in the study of physical phenomena in the complex media which is characterized by singularities in a compact space [12].

For the new algebraic operation, q-exponential and q-logarithm have the following properties:

\bea
\ln_q ( x y ) &=& \ln_q x \op \ln_q y ~~~~~ e_q (x) e_q (y) = e_q ( x \op y ) \cr
\ln_q ( x \ot y ) &=& \ln_q x + \ln_q y ~~~~~ e_q (x)\ot  e_q (y) = e_q ( x + y ) \cr
\ln_q ( x / y ) &=& \ln_q x \om \ln_q y ~~~~~ e_q (x)/  e_q (y) = e_q ( x \om y ) \cr
\ln_q ( x \od y ) &=& \ln_q x - \ln_q y ~~~~~ e_q (x)\od  e_q (y) = e_q ( x - y )
\eea

From the associativity of $\op$ and $\ot$, we have the following formula :
\be
\underbrace{t \op t \op t\op \cdots \op t }_{ n~times}= \qi \{ [\q]^n -1 \}
\ee
\be
t^{\ot^n } = \underbrace{t \ot t \ot t\ot \cdots \ot t }_{ n~times}= [n t^{1-q} - (n-1) ]^{\qi}
\ee

\section {q-Laplace transform}

In this section, we find the q-Laplace transform related to the non-extensive thermodynamics. From the relation
\be
e_q (x) e_q (y) = e_q ( x \op y ),
\ee
the q-analogue of $e^{nx}, ~(n \in Z ) $ is given by
\be
e_q ( n \odot t ) = [ e_q (t) ]^n = e_q \lb  \qi [(\q)^n -1 ] \rb = ( \q )^{ \frac{n}{1-q}}
\ee
Then we have
\be
e_q ( 0 \odot t ) =1
\ee
and the inverse of $e_q ( n \odot t )$ is $ e_q ( (-n) \odot t )$.
For this adoption, q- Laplace kernel is defined by
\be
e_q (  (-s) \odot t ) = [ e_q (t) ]^{-s} = e_q \lb  \qi [(\q)^{-s} -1 ] \rb = ( \q )^{ -\frac{s}{1-q}}
\ee

Therefore q-Laplace transform is defined by
\be
L_s (F(t)) = \int_0^{\infty } [ e_q (t) ]^{-s} F(t) d t,~~~ (s>0)
\ee
Limiting $ q \rightarrow 1 $, the eq.(13) reduces to an ordinary Laplace transform. Form now on we assume that $s$ is sufficinetly large.

Since, for two functions $F(t)$ and $ G(t)$, for which the integral exist
\be
L_s ( a F(t) + b G(t) ) = a L_s ( F(t))+ b L_s ( G(t)),
\ee
the q-Laplace transform is linear.

For $F(t) = t^N, ~(N =0, 1, 2, \cdots )   $ , we have the following result.

\vspace{1cm}

{\bf Theorem 1} \it For sufficiently large $s$ , when $q<1$, the following holds:
\be
L_s ( t^N ) =  \frac{ N!}{ (s; 1-q )_{N+1} }
\ee
where
\be
( a ; Q)_n = \cases { 1 ~~~&($ n=0 $ )\cr
\prod_{k=1}^n ( a - kQ ) ~~~&($ n \ge 1 $) }
\ee

\vspace{1cm}

Proof. \rm Let us assume that the eq.(15) holds for $t^N $. Then,
\bea
L_s ( t^{N+1} ) &=&  \int_0^{\infty } [ e_q (t) ]^{-s} t^{N+1}  d t \cr
 &=&  \int_0^{\infty } (\q  )^{ -\qs } t^{N+1}  d t \cr
 &=&  \left[ \frac{ (\q)^{-\qs +1 } }{ 1-q -s } t^{N+1} \right]_0^{\infty} + \frac{N+1}{s-(1-q)}  \int_0^{\infty } (\q  )^{ -\frac{s-(1-q)}{1-q} } t^{N}  d t \cr
  &=& \frac{N+1}{s-(1-q)} L_{s-(1-q)} ( t^{N+1} ) \cr
  &=& \frac{(N+1)N!}{(s-(1-q))(s-(1-q); 1-q)_{N+1}} \cr
    &=&  \frac{ (N+1)!}{ (s; 1-q )_{N+2} }
    \eea

In a similar way, we can obtain the q-Laplace transform for $e_q ( a \odot t ) $ which is given by
\bd
e_q ( a \odot t ) = [ e_q (t) ]^a
\ed

\vspace{1cm}

{\bf Theorem 2} \it For sufficiently large $s$ , when $q<1$, the following holds:
\be
L_s ( e_q ( a \odot t ) ) =  \frac{ 1}{ s-a - (1-q) }
\ee

\vspace{1cm}

Proof. \rm It is trivial.

In order to obtain q-Laplace transform for the trigonometric function, we need q-analogue of Euler identity. The q-Euler formula is given by
\be
e_q ( ia \odot t ) = C_q (a \odot t ) + i S_q ( a \odot t ),
\ee
where q-cosine and q-sine functions are defined by
\bd
C_q (a \odot t ) = \cos ( \frac{a}{1-q} \ln ( \q ) )
\ed
\be
S_q (a \odot t ) =  \sin ( \frac{a}{1-q} \ln ( \q ) )
\ee
and we used the following identity.
\be
p^i = e^{ i \ln p } = \cos \ln p + i \sin \ln p
\ee
Indeed, q-cosine and q-sine functions can be expressed in terms of q-exponential as follows:
\bd
C_q (a \odot t ) = \frac{1}{2} [ e_q ( ia \odot t ) + e_q ( (-ia) \odot t )
\ed
\be
S_q (a \odot t ) = \frac{1}{2i} [ e_q ( ia \odot t ) - e_q ( (-ia) \odot t )
\ee
Then we have the q-Laplace transform for q-sine and q-cosine functions :

\vspace{1cm}

{\bf Theorem 3} \it For sufficiently large $s$ , when $q<1$, the following holds:
\bd
L_s ( C_q (a \odot t ) ) =  \frac { s-(1-q)}{(s-(1-q))^2 + a^2 }
\ed
\be
L_s ( S_q (a \odot t ) ) =  \frac { a}{(s-(1-q))^2 + a^2 }
\ee

\vspace{1cm}

Proof. \rm It is trivial.

The eq.(23) can be written in terms of the ordinary sine and cosine functions :
\bd
L_s ( \cos [ \frac{a}{1-q} \ln ( \q ) ] ) = \frac { s-(1-q)}{(s-(1-q))^2 + a^2 }
\ed
\be
L_s ( \sin [ \frac{a}{1-q} \ln ( \q ) ] )=\frac { a}{(s-(1-q))^2 + a^2 }
\ee

The $q$-sine function and $q$-cosine function have the following zeros:
\be
S_q ( 1 \odot t_n ) =S_q (t_n ) =0 , ~~~ C_q ( 1 \odot u_n ) =C_q (u_n ) =0,
\ee
where
\be
t_n = \ln_q e^{n \pi } , ~~~ u_n = \ln_q e^{(n + \frac{1}{2} )\pi }, ~~~ n \in Z
\ee

From the zeros of the  $q$-sine function and $q$-cosine function, we have the following Theorem:

\vspace{1cm}

{\bf Theorem 4} \it For sufficiently large $s$ , when $q<1$, the following holds:
\be
S_q (t) = t \prod_{j=1}^{\infty} \lb 1- \frac{t}{\ln_q e^{j\pi}} \rb \lb 1- \frac{t}{\ln_q e^{-j\pi}} \rb
\ee
\be
C_q (t) = \lb 1- \frac{t}{\ln_q e^{-\pi/2}} \rb
 \prod_{j=1}^{\infty} \lb 1- \frac{t}{\ln_q e^{(j+1/2)\pi}} \rb \lb 1- \frac{t}{\ln_q e^{-(j+1/2)\pi}} \rb
\ee

\vspace{1cm}

Proof. \rm From the zeros of the $q$-sine function , we can set
\bd
\frac{ S_q (t) }{t} = A \prod_{j=1}^{\infty} \lb 1- \frac{t}{\ln_q e^{j\pi}} \rb \lb 1- \frac{t}{\ln_q e^{-j\pi}} \rb
\ed
Because $ \lim_{t \rightarrow 0 } \frac{ S_q (t) }{t} =1 $, we have $ A=1 $, which proves the eq.(27). Similarly we can easily prove the eq.(28).

Theorem 4 can be also written as follows :

\vspace{1cm}

{\bf Theorem 5 } \it For sufficiently large $s$ , when $q<1$, the following holds:
\be
S_q (t) = t \prod_{j=1}^{\infty} \left[ \q - \lb \frac { (1-q)t}{2 \sinh (\frac{j (1-q)\pi}{2} )} \rb^2 \right]
\ee
\be
C_q (t) = \lb 1- \frac{t}{\ln_q e^{-\pi/2}} \rb  \prod_{j=1}^{\infty} \left[ \q - \lb \frac { (1-q)t}{2 \sinh (\frac{1}{2}(1-q)(j+1/2)\pi ) } \rb^2 \right]
\ee

\vspace{1cm}

Proof. \rm It is trivial from the formula $ \cosh x -1 = 2 \sinh^2 \frac{x}{2} $.

\section{ q-Laplace transform and differential equation}

Now we discuss the q- differential equation. The main purpose of q-Laplace transform is in converting q-differential equation into simpler forms which may be solved more easily. Like the ordinary Laplace transform , we can compute the q-Laplace transformation of derivative by using the definition of the q-Laplace trnsform, which is given by
\be
L_s ( F'(t) ) = s L_{s+1-q} (F(t) ) - F(0)
\ee
An extension gives
\be
L_s ( F''(t) ) = s(s+1-q) L_{s+2(1-q)} (F(t) ) - sF(0)- F'(0)
\ee

Generally, we have following theorem :

 \vspace{1cm}

{\bf Theorem 6} \it For sufficiently large $s$ , when $q<1$, the following holds:
\be
L_s ( F^{(n)} (t) ) =  [s;1-q]_n L_{s+n(1-q)} (F(t) ) - \sum_{i=0}^{n-1} [s;1-q]_{n-1-i} F^{(i)} (0),
\ee
where $ F^{(0)} (0) = F(0) $ and
\be
[ a ; Q]_n = \cases { 1 ~~~&($ n=0 $ )\cr
\prod_{k=1}^n ( a + kQ ) ~~~&($ n \ge 1 $) }
\ee

\vspace{1cm}

Proof. \rm Let us assume that the eq.(34) holds for $n $. Then,
\bea
L_s ( F^{(n+1)} (t) ) &=&  \int_0^{\infty} ( \q )^{-\qs}  F^{(n+1)} (t) dt \cr
 &=&  sL_{ s + 1 -q } (   F^{(n)} (t) ) -  F^{(n)} (0)\cr
 &=&  s \{ [s+1-q;1-q]_n L_{s+(n+1)(1-q)} (F(t) ) - \sum_{i=0}^{n-1} [s+1-q;1-q]_{n-1-i} F^{(i)} (0) \}  -  F^{(n)} (0)\cr
 &=&   [s;1-q]_{n+1} L_{s+(n+1)(1-q)} (F(t) ) - \sum_{i=0}^{n} [s;1-q]_{n-i} F^{(i)} (0)
 \eea

We have another formula for the q-Laplace transform of derivative as follows:
\be
L_s ( F'(t) ) = s L_{s} \lb \frac{F(t)}{\q}  \rb - F(0)
\ee
An extension gives
\be
L_s ( F''(t) ) = s(s+1-q) L_s \lb \frac{F(t)}{(\q)^2 }  \rb - sF(0)- F'(0)
\ee
Generally, we have the following theorem:

 \vspace{1cm}

{\bf Theorem 7} \it For sufficiently large $s$ , when $q<1$, the following holds:
\be
L_s ( F^{(n)} (t) ) =  [s;1-q]_n L_{s} \lb \frac{F(t)}{(\q)^n }  \rb - \sum_{i=0}^{n-1} [s;1-q]_{n-1-i} F^{(i)} (0)
\ee

\vspace{1cm}

Proof. \rm It is not hard to prove Theorem 7.

Comparing Theorem 6 with Theorem 7, we have the following Lemma:

\vspace{1cm}

{\bf Lemma 8} \it For sufficiently large $s$ , when $q<1$, the following holds:
\be
L_s   \lb \frac{F(t)}{(\q)^n } \rb  = L_{s+n(1-q)} (F(t))
\ee

\vspace{1cm}

Proof. \rm It is trivial.

With the help of q-Laplace transform of the derivative, we can solve some differential equation. It is worth noting that $ e_q(t) $ is not invariant under the ordinary derivative, instead it obeys
\be
\frac{d}{dt} ( e_q (t) ) = \frac{1}{\q} e_q (t)
\ee
Now consider the following differential equation:
\be
F'(t)  = \frac{ F(t)}{ \q} , ~~~ F(0) =1
\ee
It is evident that $e_q(t) $ is a solution of the eq.(40).

Let us consider the vertical motion of a body in a resisting medium in which there again exists a retarding force proportional to the velocity. Let us consider that the body is projected downward with zero initial velocity $v(0) =0$ in a uniform gravitational field. The equation of motion is then given by
\be
m  \frac{d}{dt} v = m g -k v(t)
\ee
This equation is not solved by using the q-Laplace transform , instead we solve the following equation :
\be
m  \frac{d}{dt} v = m g -k \frac {v(t)}{\q}
\ee
The solution of the eq.(43) is then given by
\be
v(t) = \frac {g}{ 1- q + \frac{k}{m}} ( \q - [e_q (t)]^{-\frac{k}{m}} )
\ee

Similarly, we can modify the harmonic problem whose equation of motion is given by
\be
m \lb \frac{d}{dt}\rb^2  x(t) = - mw^2 \frac{x(t)}{(\q)^2 },
\ee
where $x(0) = A, \lb \frac{d}{dt} x \rb (0) =0 $.
The solution of the eq.(45) is then given by
\be
x(t) = A (\q)^2 \left\{ C_{\frac{3-q}{2}} \lb \sqrt{ w^2 - ( \frac{q-1}{2})^2 }\odot t )\rb +
\frac {q-1}{  \sqrt{ w^2 - ( \frac{q-1}{2})^2 } } S_{\frac{3-q}{2}} \lb \sqrt{ w^2 - ( \frac{q-1}{2})^2 } \odot t \rb \right \}
\ee

The eqs.(43)and (45) seem to be too artificial due to the factor $\q$. Instead, we can introduce the q-derivative [6] instead of the ordinary time derivative as follows:
\bd
D_t F(t) = \lim_{s \rightarrow t } \frac{ F(t) - F(s)}{ t\om s} = [\q] \frac{dF}{dt}
\ed
The Leibniz rule for q-derivative is as follows:
\be
D_t [F(t) G(t) ] = D_t [F(t)] G(t) + F(t) D_t [G(t)]
\ee
Then the eq.(43) is replaced as follows :
\be
m  D_t v = m g -k v(t)
\ee
The solution of the eq.(48) is then given by
\be
v(t) = \frac { mg}{k} ( 1 - [ e_q (t) ]^{ - \frac{k}{m} } )
\ee
Similarly, the eq.(45) is replaced as follows:
\be
m D_t^2  x(t) = - mw^2 x(t)
\ee
Using the q-Laplace transform, we get the solution of the eq.(50) :
\be
x(t) = A C_q ( w \odot t )
\ee
Obtaining these solutions, we used the following theorem:

\vspace{1cm}

{\bf Theorem 9} \it For sufficiently large $s$ , when $q<1$, the following holds:
\be
L_s ( (\q)^n F^{(n)} (t) ) =  [s-(n+1)(1-q);1-q]_n L_{s} (F(t) ) - \sum_{i=0}^{n-1} [s-(n+1)(1-q);1-q]_{n-1-i} F^{(i)} (0)
\ee

\vspace{1cm}

Proof. \rm Let us assume that the eq.(52) holds for $n $. Then,
\bea
L_s ((\q)^n  F^{(n+1)} (t) ) &=&  \int_0^{\infty} ( \q )^{-\qs +n +1 }  F^{(n+1)} (t) dt \cr
 &=&  (s-(n+1)(1-q) ) L_{ s } (  (\q)^n  F^{(n)} (t) ) -  F^{(n)} (0)\cr
 &=&  (s-(n+1)(1-q) ) \{ [s-(n+1)(1-q);1-q]_n L_{s} (F(t) ) \cr
 &-& \sum_{i=0}^{n-1} [s-(n+1)(1-q);1-q]_{n-1-i} F^{(i)} (0) \} -  F^{(n)} (0)\cr
 &=&   [s-(n+2)(1-q);1-q]_{n+1} L_{s} (F(t) ) \cr
 &-& \sum_{i=0}^{n} [s-(n+2)(1-q);1-q]_{n-i} F^{(i)} (0)
 \eea

The eq.(50) is also obtained by using the variational method whose Lagrangian is given by
\be
L = \int dt \lb \frac {1}{2} ( D_t x )^2 - U(x) \rb
\ee
The equation of motion is then given by
\be
D_t \lb \frac{ \partial L }{\partial ( D_t x ) } \rb - \frac{ \partial L }{\partial x} =0,
\ee
where the momentum $p$ is defined by
\be
p = \frac{ \partial L }{\partial ( D_t x ) }
\ee
For the harmonic potential $ U= \frac{1}{2}mw^2 x^2 $ , we have the eq.(50). This equation is rewritten by
\be
(\q)^2 \ddot{x} + (1-q) (\q) \dot{x} = - w^2 x
\ee
Replacing
\be
\eta = \frac{1}{1-q} \ln (\q),
\ee
the eq.(57) is then as follows :
\be
\frac{ \partial^2 x }{\partial \eta^2 } = -w^2 x(\eta)
\ee
Solving the eq.(59), we have
\be
x(t) = A \cos w \eta = A C_q ( w \odot t )
\ee
Here, let us investigate the times $ t_n $ when a body goes back to the initial position. This time is determined by
\be
t_n = \ln_q e^{\frac{2 \pi n }{w}},~~~(t=0,1,2,\cdots )
\ee
When $q<1$ and $w>0$, we have the following inequality:
\be
t_{n+1} - t_n > t_n - t_{n-1}
\ee
Thus, the time when a body goes back to the initial position keeps increasing.

\section{Conclusion}

In this paper, we used the algebraic operation and differential calculus related to the non-extensive thermodynamics to investigate the q-Laplace transform. We used the q-Laplace transform to solve some differential equation such as harmonic oscillator problem.

We think that this work will be applied to some q-differential equation which might appear in the study of the non-extensive statistical mechanics. We hope that these work and their related topics will be clear in the near future.

\def\JMP #1 #2 #3 {J. Math. Phys {\bf#1},\ #2 (#3)}
\def\JP #1 #2 #3 {J. Phys. A {\bf#1},\ #2 (#3)}


\section*{Refernces}

[1] C.Tsallis, J.Stat.Phys.  {\bf 52} (1988) 479.

[2] E.Curado, C.Tsallis, J.Phys. {\bf A 24} (1991) L69.

[3] A.Cho, Science {\bf 297} (2002) 1268.

[4] S. Plastino, Science {\bf 300 } (2003) 250.

[5] L.Nivanen, A.Le Mehaute , Q.Wang, Rep.Math.Phys.{\bf 52} (2003) 437

[6] E. Borges, Physica {\bf A 340} (2004) 95.

[7] S.Abe, A.Rajagopal, Science   {\bf 300 } (2003) 249.

[8] V.Latora, A,Rapisarda, A.Robledo, Science {\bf 300 } (2003) 250.

[9] F. Jackson, Mess.Math.  {\bf 38 }, 57 (1909).

[10] C.Tsallis, Quimica Nova  {\bf 17} (1988) 479.

[11] A.Beardon, An Introduction to hyperbolic geometry, Oxford University Press, New York, (1991).

[12] A.Mehaute, R.Nigmatullin, L.Nivanen, Fleches du temps et geometrie fractale. Hermes, Paris, (1998).

\end{document}